\documentclass[aps,physrev,twocolumn,superscriptaddress]{revtex4-2}

\usepackage{graphicx}
\usepackage{xcolor}
\usepackage[colorlinks, linkcolor=mycol, citecolor=mycol, urlcolor=mycol, breaklinks]{hyperref}
\usepackage{bbm}
\usepackage{amsmath}
\usepackage{braket}

\definecolor{mycol}{RGB}{10,55,130}

\begin{document}

\title{Efficiency of classical simulations of a noisy Grover algorithm}

\author{R.~Menu}
\affiliation{CESQ/ISIS (UMR 7006), CNRS and Universit\'{e} de Strasbourg, 67000 Strasbourg, France}

\author{J.~Schachenmayer}
\affiliation{CESQ/ISIS (UMR 7006), CNRS and Universit\'{e} de Strasbourg, 67000 Strasbourg, France}

\date{\today}

\begin{abstract}
We analyze the modification of entanglement dynamics in the Grover algorithm when the qubits are subjected to single-qubit amplitude-damping or phase-flip noise. We compare quantum trajectories with full density-matrix simulations, analyzing the dynamics of averaged trajectory entanglement (TE) and operator entanglement (OE), in the respective state representation. Although not a genuine entanglement measure, both TE and OE are connected to the efficiency of matrix product state simulations and thus of fundamental interest. As in many quantum algorithms, at the end of the Grover circuit entanglement decreases as the system converges towards the target product state. While we find that this is well captured in the OE dynamics, quantum trajectories rarely follow paths of reducing entanglement. Optimized unraveling schemes can lower TE slightly, however we show that deep in the circuit OE is generally smaller than TE. This implies that matrix product density operator (MPDO) simulations of quantum circuits can in general be more efficient than quantum trajectories. In addition, we investigate the noise-rate scaling of success probabilities for both amplitude-damping and phase-flip noise in Grover's algorithm.
\end{abstract}

\maketitle

\section{Introduction}


Entanglement is a key resource for achieving a quantum advantage in practical quantum algorithms. This connection can be made with concept of matrix product states (MPS)~\cite{Vidal_2004, Scholwoeck_2011}, also known as tensor trains in other communities~\cite{Oseledets_2011}. An MPS can provide an efficient classical representation of a quantum state as long as the bipartite entanglement entropy remains small~\cite{Schuch_2008}. The temporary presence of large-scale entanglement between blocks of qubits in a quantum register is thus a necessary condition for achieving a quantum advantage. In recent quantum computing platforms qubits remain subjected to noise, and it has been controversially debated whether such platforms can lead to a practical quantum advantage~\cite{Babbush_2021, Hoefler_2023, stoudenmire_2024}. In this context it has become an important task to understand the impact of noise on the evolution of entanglement.

Characterizing bipartite entanglement between blocks of qubits in an open quantum circuit is however a challenging problem, as finding genuine quantum entanglement for large systems, such as entanglement of formation~\cite{Wooters_1998}, is an NP-hard problem~\cite{Gurvits_2003, Ghariaban_2010}. Following the unraveling of open-system dynamics into pure-state trajectories~\cite{Molmer_93, Dum_1992, Gardiner_1992, Daley_2014}, one can evaluate the average bipartite entanglement entropy for individual trajectories and define a quantity dubbed as trajectory entanglement (TE). Though this quantity is no measure of non-separability, it provides an assessment of the efficiency of MPS simulations. However, TE is not uniquely defined and can differ widely depending on the chosen unraveling strategy. This freedom results from the infinite manners to decompose a density matrix into a product of pure states, and is exploited to minimize entanglement and maximize the efficiency of MPS simulations via, for instance, an adjustment of the unraveling scheme~\cite{Vovk_2022,kolodrubetz2023optimality}. These approaches are however computationally challenging and spurred the search for feasible adaptative schemes~\cite{cheng2023efficient,Vovk_2024,Daraban_2025, cichy2025classical, chen2024optimized}.

\smallskip

In this work, we investigate the effect of typical types of single-qubit noise processes, encountered on experimental quantum computing platforms, namely phase-flip and amplitude-damping channels. We analyze their impact on classical simulability by comparing their influence on the growth of trajectory entanglement for Grover's algorithm \cite{Grover_1996}. The Grover search algorithm is believed to hold the potential for a quadratic speedup compared to classical unstructured search, as long as its black-box function dubbed ``quantum oracle'' generates large amounts of entanglement~\cite{stoudenmire_2024}. Keeping the oracle black box closed, here our goal is to analyze the additional impact of noise processes in the Grover circuit, to shed light on the acceptable noise rates for efficient classical simulations, either with MPS decompositions of quantum trajectory states, or with a matrix product density operator representation (MPDO)~\cite{zwolak2004mixed,GuthJarkovsky_Effici_2020,weimer2021simulation}.

\smallskip

We find that even for small noise rates, MPDO simulations tend to be more efficient than trajectory unravelings. We furthermore determine the scaling behavior of success probabilities in Grover's algorithm as function of the noise rates and qubit number. While, as expected, the success probabilities vanish exponentially as function of the qubit number, we find generally an algebraic decay as function of the noise rate. This paper is structured as follows: after reviewing the mathematical framework underlying Grover's algorithm in Sec.~\ref{sec:grover}, we analyze the effect of noise in quantum trajectories and in the MPDO approach in Sec.~\ref{sec:ent}. In Sec.~\ref{sec:scale}, we assess the performances of Grover's algorithm as a function of the qubit number and the noise rate.  We provide a summarizing conclusion and an outlook in Sec.~\ref{sec:concl}.

\section{Grover search algorithm}
\label{sec:grover}

\begin{figure}
    \centering
    \includegraphics[width=\columnwidth]{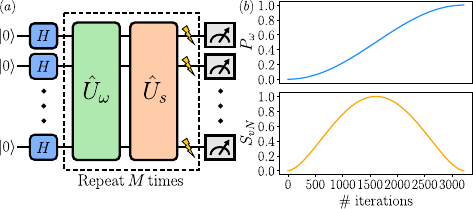}
    \caption{Grover search algorithm - (a) Representation of the circuit implementing the Grover search algorithm. In the following we will assume that after each Grover iteration $\hat{U}_\omega\hat{U}_s$ a single-qubit noise channel is applied on each qubit. (b) Evolution of the success probability $P_\omega$ and the von Neumann entropy $S_{vN}$ for an equal bipartition of the qubit string as a function of the number of iterations of the Grover search. Calculations are performed for $n=24$ qubits.}
    \label{fig:1}
\end{figure}

The search of a single entry in an unstructured database of $N$ elements is realized via quantum resources using the Grover algorithm \cite{Grover_1996}. To this end, let us consider a string of $n$ qubits (described by the local basis $\lbrace\vert 0 \rangle, \,\vert 1 \rangle\rbrace$), such that the database is mapped onto the $N=2^n$ elements of the qubits' Hilbert space. One entry of this database --or conversely a state of the qubits' Hilbert space-- is marked as the target of the search, encoded by the state $\vert \omega\rangle$. The outline of the Grover search algorithm in its quantum-circuit implementation is depicted in Fig.~\ref{fig:1}(a): first all qubits are initialized in state $\vert 0 \rangle$ before entering into a set of Hadamard gates. The resulting state reads 
\begin{equation}
    \vert\psi_0\rangle = \vert s \rangle=\bigotimes_{j=1}^n\vert + \rangle_j=\frac{1}{\sqrt{N}}\sum_{i=1}^N\vert i \rangle,
\end{equation}
namely a coherent superposition of the all states $\vert i \rangle$ of the Hilbert space with equal weights $1/N$. Then, a global gate is applied to all qubits at once that implements the oracle operator $\hat{U}_\omega$, a black box that recognizes the target state $\vert\omega\rangle$ and inverts its phase while leaving all other states invariant. Formally, the effect of the oracle is represented by the operator
\begin{equation}
    \hat{U}_\omega = \mathbbm{1} - 2\vert\omega\rangle\langle\omega\vert.
\end{equation}
The action of the oracle is followed by the diffusion operator, which inverts the state with respect to the uniform initial state $\vert s \rangle$: 
\begin{equation}
    \hat{U}_s = 2\vert s \rangle\langle s \vert - \mathbbm{1},
\end{equation}
The oracle and diffusion operators are sequentially applied to the circuit's state, such that at the $k$-th iteration it reads $\vert\psi_k\rangle =  \hat{G}^k\vert\psi_0\rangle = (\hat{U}_s \hat{U}_\omega)^k\vert \psi_0\rangle$. The success probability $P_\omega =\vert\langle\omega\vert\psi\rangle\vert^2$ reaches a maximum close to unity after an optimal number $M$ of iterations that scales with the size $N$ of the Hilbert space:
\begin{equation}
    M\simeq \frac{\pi}{4}\sqrt{N},
\end{equation}
which display a quadratic speed up compared to a classical search whose computation complexity scales as $O(N)$. The evolution of the success probability, $P_\omega$, as a function of the number of Grover iterations is displayed in Fig.~\ref{fig:1}(b). There we also plot the evolution of the von Neumann entropy $S_{vN}=\mathrm{Tr}[-\hat{\rho}_{A/B}\log_2 \hat{\rho}_{A/B}]$, where $\hat{\rho}_{A/B}$ stands for the reduced density matrix for a bipartition of the chain of qubits into two equal blocks of qubits $A=\{1,\dots,n/2\}$ and $B=\{n/2+1,\dots,n\}$. Remarkably, independently from the number of qubits $n$, the amount of entanglement generated over a realization of the Grover search remains weak and displays an upper bound at unity, which follows from an effective two-level description of the algorithm (see Appendix). This very low bound on the amount of entanglement and its independence on the number of qubits suggests that the evolution of the circuit's state during a Grover search can be exactly simulated by the means of a matrix product state (MPS), unless entanglement is built-up within the black box~\cite{stoudenmire_2024}.

\section{Trajectory and operator entanglement}
\label{sec:ent}

\begin{figure*}
    \centering
    \includegraphics[width=0.8\textwidth]{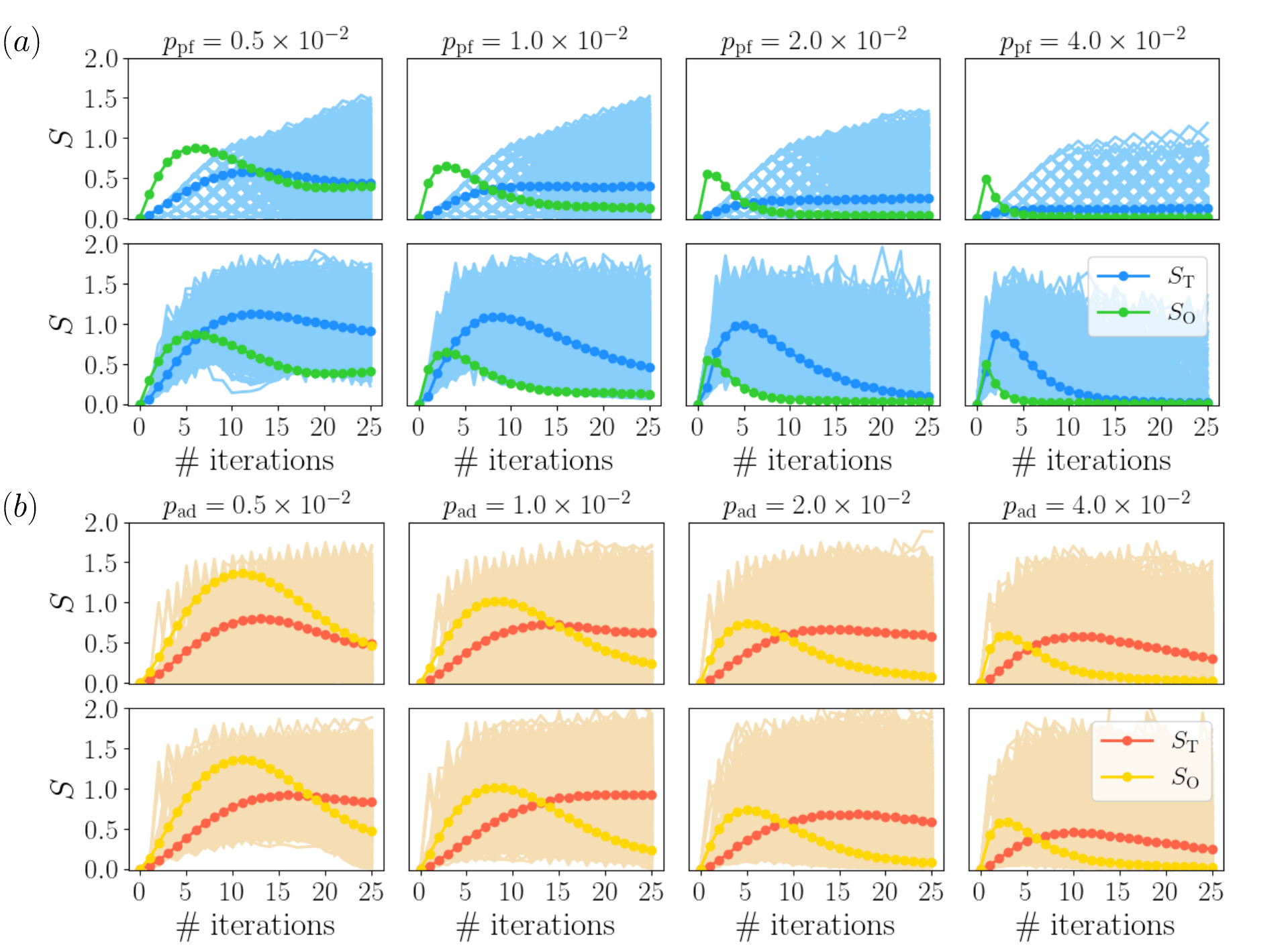}
    \caption{TE and OE evolution in the Grover search algorithm subjected to (a) phase flip, and (b) amplitude noise channels ($n=10$ qubits). In each panel we plot the von Neumann entropy of $n_T=2000$ individual trajectories unraveled via a a naive (first row) or NUMU (second row) procedure. Different columns show results for different noise rates $p_{\rm pf/ad} = 0.5 \times 10^{-2}$, $1.0 \times 10^{-2}$, $2.0 \times 10^{-2}$, and $4.0 \times 10^{-2}$. We compare the averaged TE, $S_\mathrm{T}$,  to the OE, $S_\mathrm{OE}$. The  entropies are relevant to the efficiency of matrix product decompositions of the trajectory states, or the full density matrix, respectively.}
    \label{fig:2}
\end{figure*}

A main limitation for achieving a useful quantum advantage is the inescapable effect of couplings to the environment. In Grover's algorithm dissipation and dephasing lead to errors that manifest as a decrease of the success probability. In terms of classical simulations, adding noise seems to make the simulations more difficult, due to the shift from a pure state vector to a density matrix representation. However, the small amount of entanglement, even in the noiseless case, suggests an MPS representation. As noise is expected to lower entanglement, such an MPS representation should then become even more efficient in the presence of noise. Here, we discuss the simulability of a noisy Grover circuit using two distinct MPS approaches.

\smallskip

We describe single-qubit noise by the quantum operation formalism, i.e.~by a general linear map of the density matrix in the operator sum representation~\cite{nielsen2010quantum}, with Kraus operators:
\begin{equation}
    \hat{\rho}' = \mathcal{E}(\hat{\rho}) = \sum_n {\hat{E}_n \hat{\rho} \hat{E}^\dagger_n}.
\end{equation}
Here, the  Kraus operators $\lbrace\hat{E}_n\rbrace$ must respect the constraint $\sum_n{\hat{E}^\dagger_n \hat{E}_n} = \mathbbm{1}$. Typically, for large system sizes, it is less numerically costly to unravel the dynamical map via the framework of quantum trajectories: it is easy to see that dynamics of the open-system density matrix can be approximated via an averaging over a large number $n_t$ of pure states
\begin{equation}
    \hat{\rho} = \lim_{n_t \to +\infty} \frac{1}{n_{t}}\sum_{k=1}^{n_t}\vert \psi^{(k)}\rangle\langle\psi^{(k)}\vert,
\end{equation}
where individual trajectories $\vert \psi^{(k)}\rangle$ are generated stochastically by applying Kraus operators $\vert\psi'^{(k)}\rangle = \hat{E}_j\vert\psi^{(k)}\rangle/\sqrt{p_j}$ with probability $p_j = \langle \psi^{(k)} \vert \hat{E}^\dagger_j \hat{E}_j \vert\psi^{(k)}\rangle$.

Each trajectory is associated with a given amount of entanglement that can be measured by the von Neumann entropy of a reduced density matrix $\hat{\rho}^{(k)}_A$, such that $S^{(k)}=-\mathrm{Tr}[\hat{\rho}^{(k)}_A \log_2 \hat{\rho}^{(k)}_A]$. The average of the entanglement entropy of individual pure states defines a trajectory entanglement (TE)
\begin{equation}
    S_T = \frac{1}{n_t}\sum_{k=1}^{n_t}{S^{(k)}}.
\end{equation}
We note that TE is distinct from the von Neumann entropy of reduced density matrices computed from the full density matrix $\hat{\rho}$, the operator entanglement (defined below), and the genuine entanglement of formation~\cite{Wooters_1998}. Nevertheless, it is a relevant quantity for this work, as it is connected to the necessary ressources needed on average to simulate an open quantum circuit by the means of a trajectory MPS representation.

\smallskip

Importantly, TE is a highly non-unique quantity and depends on the choice of Kraus operators. Indeed, Kraus operators are not uniquely defined: With any unitary transformation, one can define an equivalent set $\lbrace \hat{F}_m \rbrace$ of Kraus operators that generate the exact same dynamical map as the operators $\lbrace \hat{E}_n\rbrace$, namely $\hat{F}_m = \sum_n \mathcal{U}_{mn}\hat{E}_n$. Despite describing upon averaging the same expectation values for observables, individual trajectories computed for several choices of Kraus operators may appear very different, with large variations in TE~\cite{Vovk_2024, Daraban_2025}. As a result, by choosing adequately the unitary transformation $\mathcal{U}$, one may attempt to minimize TE and thus enhance the MPS simulability. Such strategies are however typically computationally costly, which e.g.~motivated the development of a method coined as non-unitarity maximizing unraveling (NUMU)~\cite{Daraban_2025}. By the means of a quantity dubbed as trajectory non-unitarity, whose calculation is much simpler than the von Neumann entropy, this approach proposes a computationally cheap adaptive optimization scheme for lowering TE.

In this work, we not only compare different types of unraveling schemes --a naive one and NUMU--, but also a distinct tensor-network approach that does not require any unraveling. To do so, we describe the full system density matrix $\hat \rho$ by a matrix-product density operator (MPDO)~\cite{zwolak2004mixed,GuthJarkovsky_Effici_2020,weimer2021simulation}. Fundamentally, MPDOs have the same architecture as MPS representations and similar limitations in terms of amount of correlations that can be captured. For an MPDO, one may measure the amount of correlations between bipartitions using an operator entanglement (OE)~\cite{zanardi2000entangling,zanardi2001entanglement,wang2002quantum,prosen2007operator, dubail2017entanglement}. OE follows from an Schmidt decomposition of $\hat \rho$ into two blocks $\hat{\rho} = \sum_a \lambda_a \hat{\rho}^{(L)}_a \otimes \hat{\rho}^{(R)}_a$, where $\lambda_a$. The, OE is defined as
\begin{equation}
    S_O = -\sum_{a}{\lambda_a^2 \log_2 \lambda_a^2}.
\end{equation}
While not being a genuine entanglement measure, OE quantifies the faithfulness of the MPDO representation to the actual many-body dynamics of a system~\cite{Wellnitz_2022, Preisser_2023}.

\medskip

A matrix product decomposition --both for trajectory states and the MPDO-- is a numerical decomposition into a product of $\chi \times \chi$ matrices, where the entries of the matrices are local states~\cite{Scholwoeck_2011, weimer2021simulation}. In the pure-state case the local entries are state vectors of dimension $2$, while for an MPDO, local entries are single-qubit density matrices requiring a local dimension of $4$. The matrix dimension $\chi$ is commonly known as bond dimension, and naturally connected to the amount of bipartite entropy. For example, it is easy to see that in such a representation, the bipartite entropy is fundamentally limited to $\max[S_{T/O}] = \log_2(\chi)$ for any bipartition. Updates of matrix product decompositions can be achieved via gate applications using a TEBD algorithm or via DMRG-like sweeps~\cite{Scholwoeck_2011, weimer2021simulation}, leading in both cases to a numerical complexity $O (\chi^3)$.

\medskip

For Grover's algorithm, we compare TE and OE dynamics in Fig.~\ref{fig:2}. Considering an equal bipartition of the qubit chain, we compute the von-Neumann entropy for individual MPS trajectories whose large deviations lead to an effective shaded area in Fig.~\ref{fig:2}\,(a) for a phase-flip noise described by the single-channel Kraus operators
\begin{equation}
    \hat{E}^{\rm pf}_1 = \sqrt{p_{\rm pf}}\,\hat{\sigma}^z~;~\hat{E}_2^{\rm pf} = \sqrt{1-p_{\rm pf}}\,\hat{\mathbbm{1}}.
\end{equation}
Fig.~\ref{fig:2}\,(b) represents the behavior of individual-trajectory entanglement in the case of amplitude damping, with
\begin{equation}
    \hat{E}^{\rm ad}_1 = \sqrt{p_{\rm ad}}\,\hat{\sigma}^- ~;~\hat{E}^{\rm ad}_2 = \vert 0 \rangle\langle 0 \vert + \sqrt{1-p_{\rm ad}}\,\vert 1 \rangle\langle 1 \vert.
\end{equation}
In both cases, we apply the noise channels to each qubit after each iteraction, i.e.~after each application of $\hat U_\omega \hat U_s$, as sketched in Fig.~\ref{fig:1}(a). The noise rate is given by $p_\mathrm{pf/ad}$. In the case of amplitude damping, $p_\mathrm{ad}$ corresponds to the probability of applying a quantum jump $\ket{1} \to \ket{0}$ (via $\hat \sigma_-$) within a given time window, while, equivalently for phase-flip, $p_\mathrm{pf}$ corresponds to a spontaneous $\pi$-phase addition to the state $\ket{0}$ (via $\hat \sigma_z$).  We compare a naive unraveling scheme with the bare Kraus operators $\hat{E}_{1,2}$ with the NUMU procedure~\cite{Daraban_2025}, leading to distinct TE ($S_T$). We also compare the TE with the OE ($S_O$) for an equivalent exact simulation of the full density matrix dynamics. 

Regardless of its nature, we observe that the noise channels generate individual trajectories that depart from the effective two-level description of the Grover search algorithm (see Appendix). They lead to a high amount of entanglement that overshoots the upper bound at unity. We deduce that the faithful description of the noisy dynamics of the Grover circuit is only achievable with MPS bond dimensions $\chi >2$. Such trajectories hold more statistical significance for weak values of $p_{\rm pf/ad}$ as it can be deduced from the dynamics of $S_T$. They correspond to rarer events for large $p_{\rm pf/ad}$ as hinted by the decrease of $S_\mathrm{T}$. We also notice that the optimized NUMU approach  only reduces $S_\mathrm{T}$ by a small amount. We conclude that the local adaptive optimization of entanglement does not seem to be sufficient to lead to a significant TE reduction compared to the naive unraveling.

\smallskip

In contrast, OE displays consistently smaller values, always deep in the circuit, and even more so for large $p_{\rm pf}$ and $p_{\rm ad}$. This indicates that in most cases the states appearing in the Grover circuit can be better described by an MPDO representation with bond dimension of even just $\chi = 2$, a larger amount of resources only being needed for weak amplitude-damping noise. Given that the numerical complexity for MPDO updates is limited by the bond dimension, scaling as $\chi^3$, despite requiring a larger local dimensions, an MPDO can thus fundamentally outperform MPS-based unravelings, independently of the choice made for the Kraus operators. 

\smallskip

In the noiseless Grover algorithm, towards the end of the circuit, entanglement decreases as the state converges towards $\ket{\omega}$, a product state. This behavior is a generic feature of many useful quantum algorithms, since preparing a highly entangled state would require many re-runs of the circuit in order to extract useful information from projective measurements of the final state. With the addition of noise, when using quantum trajectories, here we found that it can be very difficult to find an unraveling strategy that captures the behavior of decreasing TE. Random Kraus operator applications lead to trajectory states that are far from the ``noiseless'' trajectory with decreasing entanglement. Instead, on the density-matrix level, our results suggest that one can still observe a decreasing OE towards the end of the circuit. The counter-intuitive fact that a simulation of the full density matrix (in MPDO form) can be fundamentally numerically more efficient than pure state trajectory unravelings could therefore hold for several other quantum algorithms.

\section{Noise rate scaling}
\label{sec:scale}

\medskip
\begin{figure*}
    \centering
    \includegraphics[width=0.85\textwidth]{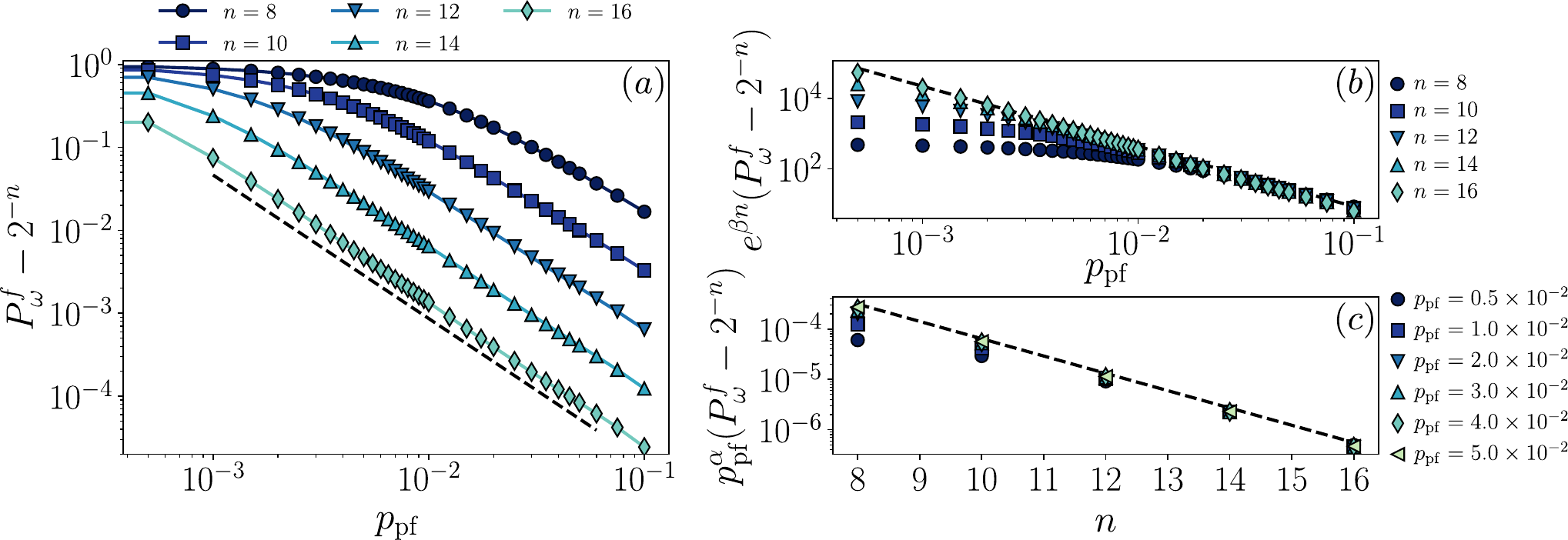}
    \caption{(a) Final success probability $P_\omega^f$ after $M = (\pi/4) \sqrt{N}$ iterations as a function of the phase-flip rate $p_{\rm pf}$ for various number of qubits $8 \leq n \leq 16$ (log-log scale). The dashed black line stands for an algebraically decaying function of the form $p_{\rm pf}^{-\alpha}$ where $\alpha = 1.735$. Results are obtained for an exact MPDO representation (converged with $\chi = 4$). (b)-(c) Scaling analysis of the final success probability with respect to $p_{\rm pf}$ (upper panel) and number of qubits $n$ (lower panel). Dashed lines indicate the algebraic ($P^f_\omega \propto p_{\rm pf}^{-\alpha}$) and exponential ($P^f_\omega \propto e^{-\beta n}$) dependence, respectively.
    \label{fig:3}}
\end{figure*}

Previous analyses of the applicability of Grover's algorithm in an open environment have mainly focused on idealized forms of noise that act in the effective two-dimensional Hilbert space \cite{pablo1999noise, shenvi2003effects}. In a recent work it was argued that a noise channel can lead to a super-exponential decrease with qubit number $n$~\cite{stoudenmire_2024}. This is true for the specific case of a global depolarizing noise described by the map $\mathcal{E}(\hat\rho) = (1-p)\hat{\rho} + (p/2^n) \hat{\mathbbm{1}}$. Then, it is straightforward to see that the final success probability $P^f_\omega$, after $M=(\pi/4)\sqrt{N}$ iterations (rounded down to the nearest integer), follows the exponential scaling 
\begin{equation}
    P_\omega^f = \exp(-pM).
\end{equation}

\smallskip

Here, our goal is to analyze the scaling of $P_\omega^f$ as function of qubit numbers $n$ and noise rates $p_\mathrm{pf/ad}$ for the cases of phase flip and amplitude damping, as defined in Sec.~\ref{sec:ent}.

\medskip

\begin{figure*}
    \centering
    \includegraphics[width=0.85\textwidth]{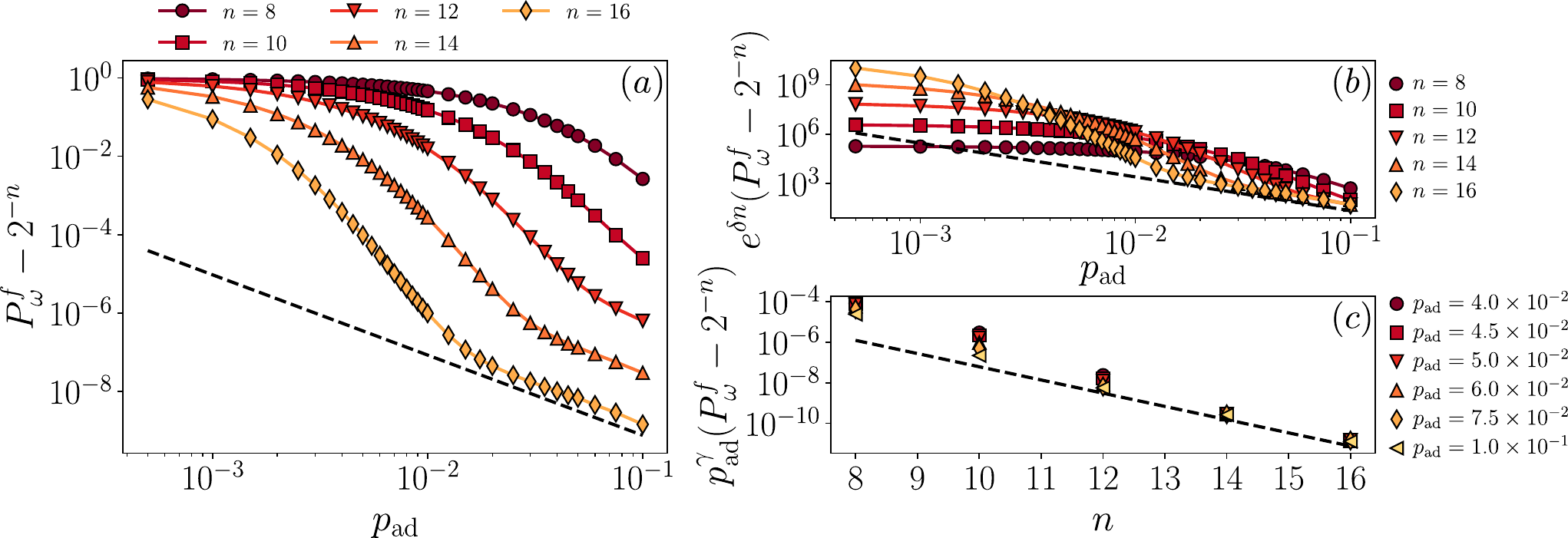}
    \caption{(a) Final success probability $P_\omega^f$ shifted by $2^{-n}$ after $M = (\pi/4) \sqrt{N}$ iterations as a function of the amplitude-damping rate $p_{\rm ad}$ for various numbers of qubits $8 \leq n \leq 16$ (log-log scale). The dashed black line stands for an algebraically decaying function of the form $p_{\rm pf}^{-\alpha}$ where $\alpha = 1.7$. Results are obtained for an exact MPDO representation (converged with $\chi = 4$). (b)-(c) Scaling analysis of the shifted final success probability with respect to $p_{\rm ad}$ (upper panel) and number of qubits $n$ (lower panel). Dashed lines indicate the algebraic ($P^f_\omega - 2^{-n} \propto p_{\rm ad}^{-\gamma}$) and exponential ($P^f_\omega - 2^{-n} \propto e^{-\delta n}$) dependence, respectively.}
    \label{fig:4}
\end{figure*}

We first consider the effect of the dephasing mechanism given by the Kraus operators $\hat E^{\rm pf}_{1,2}$, applied independently on every qubit after each Grover iteration. The evolution of the success probability $P_\omega$ is extracted using an exact MPDO representation of the circuit state. The final value $P^f_\omega$ of the success probability is plotted in Fig.~\ref{fig:3} both as a function of the probability $p_\mathrm{pf}$ and the qubit number $n$. In accordance with what is expected, the final success probability decreases as both $p_\mathrm{pf}$ and $n$ increase. However, the scaling of the success probability with $p_\mathrm{pf}$ differs strongly from the one observed for a global depolarizing channel in~\cite{stoudenmire_2024}. We identify an algebraic dependence on $p_\mathrm{pf}$ [log-log scale in Fig~\ref{fig:3}(a)], and a simple exponential dependence as function of $n$, such that we may fit the scaling function: 
\begin{equation}
    P^f_\omega \simeq \dfrac{1}{2^n}+ \frac{e^{-\beta n}}{p_\mathrm{pf}^\alpha}.
\end{equation}
Numerically we can fit parameters $\alpha = 1.735\pm0.007$ and $\beta = 0.7844\pm 0.001$. This implies that Grover's algorithm is  remarkably robust against the case of a local phase-flip noise as it displays a scaling of the success probability that is much more advantageous in terms of the noise rate $p_\mathrm{pf}$ and as a function of the qubits' number $n$ compared to global depolarizing noise. We note that for all phase-flip rates up to $p_{\rm pf} = 0.1$ the success probability remains above the probability of finding a random bitstring. The latter is determined by the Hilbert space size to be $2^{-n}$, e.g.~for $n=16$, $2^{-16}\approx 1.5 \times 10^{-5}$. In Fig.~\ref{fig:3} and~\ref{fig:4} we subtract $2^{-n}$ in order to only focus on the ``excess probability'' above the trivial random bitstring probability.

\medskip

Secondly, we discuss the success probability for amplitude damping with the Kraus operators $\hat{E}^{\rm ad}_{1,2}$, applied independently on every qubit after each Grover iteration. Importantly, while phase-flip noise is fundamentally independent on the choice of target state $\vert \omega \rangle$, this is not true for amplitude damping. As this channel is spontaneously flipping qubit states, the success probability over the number of Grover iterations depends on the number $n_1$ of states $\vert 1 \rangle$ in the qubit sequence of the target state $\vert \omega \rangle$. Therefore, for a string of $n$ qubits, we expect to find $n+1$ possible values of the success probability depending on $n_1$. We compute the final success probability $P^f_\omega$ as the statistical average over all possible choices of target states
\begin{equation}
    P^f_\omega = \sum_{n_1=0}^n{p_{n_1} P^f_{n_1}},
\end{equation}
where $P^f_{n_1}$ is the success probability obtained for a target state with given $n_1$, and $p_m = C^n_m/2^n$ with $C^n_m = \binom{n}{m} =  \frac{n !}{m! (n-m)!}$ the binomial coefficients.  The scaling of success probabilities is shown in Fig.~\ref{fig:4} as a function of $p_{\rm ad}$ and $n$. As in the case of phase flip, naturally an increase of $p_{\rm ad}$ and $n$ leads to a decrease of $P^f_\omega$. However, we observe that the Grover search is much more sensitive to amplitude damping, as $P^f_\omega$ decays rapidly towards the asymptotic random bitstring value value of $2^{-n}$.

As for the phase-flip noise, we investigate the scaling of the success probability as it converges towards its asymptotic value, whose behavior is displayed on Fig.~\ref{fig:4}. It is readily noticeable that asymptotically the success probability for amplitude damping exhibits a form that is very similar to the one already observed in the phase-flip case, namely the combination of an algebraic and exponential decay law,
\begin{equation}
    P_\omega^f \simeq \frac{1}{2^n} + \frac{e^{-\delta n}}{p_\mathrm{ad}^\gamma}.
\end{equation}
From the scaling analysis shown on Fig.~\ref{fig:4}  we deduce that the exponents $\gamma,\delta$ differ from the one found in the case of phase-flip noise as a fitting gives $\gamma = 2.050 \pm 0.034$ and $\delta = 1.522\pm0.004$. Although not as advantageous as the case of dephasing noise, we still observe that a dissipative Grover circuit does not display the super-exponential decay with $n$ predicted for a depolarizing type of noise.

However, it is important to point out that the universal scaling is only observable asymptotically for large $n$ and large $p_{\rm ad}$, in regimes where the success probabilities above the trivial value $2^{-n}$ are extremely small ($\lesssim 10^{-6}$). In regimes, where the Grover search features a reasonably high success chance, $P^f_\omega$ features a non-universal exponential decrease in both $n$ and $p_{\rm pf}$.

\section{Conclusion \& Outlook}
\label{sec:concl}

In this work, we have studied the effect of single-qubit noise, in the form of phase-flip and amplitude-damping channels, on the Grover search algorithm. We have analyzed how the noise impacts the evolution of trajectory entanglement (TE) in various unraveling schemes and compared it to the evolution of operator entanglement (OE) in the density matrix. Remarkably, we find that OE can be significantly lower than TE. This implies that numerical simulations using matrix product density operator (MPDO) representations can be in practice more efficient than simulating the algorithm with trajectories on the state-vector level. This can be rationalized by the fact that the entanglement in the ideal Grover algorithm is only temporarily present, and removed from the register as the algorithm converges to the final product state. Then, for a stochastic unraveling scheme it becomes very difficult to follow a trajectory that lowers entanglement with circuit depth. In contrast, a density matrix can still more efficiently capture this feature. As the temporary presence of entanglement is a generic feature of many quantum algorithms, we expect this conclusion to also hold for such other algorithms.

Furthermore, we analyzed the scaling of success probabilities of Grover's algorithm under the single-qubit noise channels. We find that asymptotically for large noise and large qubit numbers, the success probability only decreases algebraically with the respective noise rates, and exponentially with the qubit number. This is in contrast to the super-exponential decay with qubit number for a global depolarizing channel. 

In the future it would be interesting to contrast TE and OE evolution also in other quantum algorithms subjected to noise. In particular, the hypothesis that disentangling dynamics towards the end of an algorithm is always more efficiently captured in the OE compared to the TE would be interesting to test in a broader context. In addition, since the unraveling achieved by NUMU is far from optimal, it is a promising prospect to use Grover's algorithm for finding more advanced adaptive unraveling optimization schemes.

\appendix
\section{Upper bound on entanglement entropy in the Grover search algorithm}

Over the course of the Grover search algorithm, the state $\vert \psi_k \rangle$ of the circuit of the $k$-th iteration can be written in a compact way as 
\begin{equation}
    \vert\psi_k\rangle = \hat{G}^k\vert \psi_0\rangle = \sin\theta_k\vert \omega\rangle + \cos\theta_k\vert r \rangle,
\end{equation}
where $\vert\omega\rangle$ is the target state while $\vert r \rangle = \frac{1}{\sqrt{N-1}}\sum_{i\neq\omega}\vert i \rangle$ is a superposition of all other states of the Hilbert space \cite{Rossi_2013}. The corresponding density matrix takes then the form
\begin{align}
    \hat{\rho}_k &= \sin^2\theta_k\vert \omega\rangle\langle\omega\vert + \cos^2\theta_k\vert r \rangle\langle r \vert \notag\\
    &+ \sin\theta_k\cos\theta_k(\vert\omega\rangle\langle r\vert + \vert r \rangle\langle\omega\vert).
\end{align}
Written in the basis $\lbrace \vert\omega \rangle, \vert i \rangle_{i\neq \omega} \rbrace$, the density matrix exhibits a peculiar form:
\begin{equation}
    \hat{\rho}_k = \begin{pmatrix} a & b &\cdots& b \\
    b & c &\cdots &c \\
    \vdots & \vdots &\ddots & \vdots\\
    b & c &\cdots  &c
    \end{pmatrix},
\end{equation}
where $a = \sin^2\theta_k$, $b = \frac{1}{\sqrt{N-1}}\sin\theta_k\cos\theta_k$ and $c=\frac{1}{N-1}\cos^2\theta_k$. Taking an equal bipartition of the qubits' chain that generates the register of states $\lbrace\vert i \rangle\rbrace$, one applies a partial trace to density matrix $\hat{\rho}_k$ such that the reduced density matrix for an equal bipartition of the chain of qubits reads:
\begin{equation}
    \hat{\rho}^{(1/2)}_k = \begin{pmatrix} \alpha & \beta &\cdots& \beta \\
    \beta & \gamma &\cdots & \gamma \\
    \vdots & \vdots &\ddots & \vdots\\
    \beta & \gamma &\cdots  & \gamma
    \end{pmatrix},
\end{equation}
whose matrix elements take the form
\begin{align*}
        \alpha &= \left(\frac{\sqrt{N}}{N-1} - \frac{1}{N-1}\right)\cos^2\theta_k + \sin^2\theta_k\\
        \beta &= \left(\frac{\sqrt{N}}{N-1} - \frac{1}{N-1}\right)\cos^2\theta_k + \frac{1}{\sqrt{N-1}}\cos\theta_k\sin\theta_k\\
        \gamma &= \frac{\sqrt{N}}{N-1}\cos^2\theta_k.
\end{align*}
The reduced density matrix can be readily diagonalized: all of its eigenvalues vanish at the exception of only two that we note $\lambda_\pm$ and take the following form
\begin{equation}
    \lambda_\pm = \frac{1}{2} \pm\frac{1}{2}\left[(\alpha - (\sqrt{N}-1)\gamma)^2 + 4(\sqrt{N}-1)\beta^2\right]^{1/2}.
\end{equation}

We deduce then then that the Schmidt rank for a bipartition of the chain is exactly equal to $2$, which bounds entanglement entropy from above by the value $\max S_{vN} = \ln 2$ independently from the size of the register $N$. The maximum of the von Neumann entropy expressed as $S_{vN} = -\mathrm{Tr}[\hat{\rho}^{(1/2)}\log_2 \hat{\rho}^{(1/2)}] = -\lambda_+\log_2 \lambda_+ - \lambda_-\log_2\lambda_-$ is reached for $\cos\theta_k = \sin\theta_k = \frac{\sqrt{2}}{2}$. This maximum still depends on $N$, however in the limit $N\to\infty$, both eigenvalues take the form $\lim_{N\to \infty}\lambda_\pm = 1/2$, which confirms that the upper bound is reached $\lim_{N\to \infty}S_{vN}=1$.

\section*{Acknowledgments}

The authors thank G. Morigi and R. Daraban for their insightful remarks. This work was funded by the ERC Consolidator project MATHLOCCA (Grant nr.~101170485), and by the French National Research Agency under the France 2030 program ANR-23-PETQ-0002 (PEPR project QUTISYM) as well as the Investments of the Future Program project ANR-21-ESRE-0032 (aQCess).

\vspace{1cm}

\bibliography{main.bib}

\end{document}